\documentclass{PoS}
\PoS{PoS(LAT2005)317}
\usepackage{epsfig}
\newcommand{\prepri}[1]{
\author{
\flushright
\begin{minipage}[t]{2.8cm}{\vspace{-1cm} \rm #1\vfill\vspace{0.5cm}} 
\end{minipage}
	}
}

\title{
  Parameter dependence of the topology change 
  and the scaling properties of the topology conserving
  gauge action
}
\ShortTitle{Topology conserving gauge action }

\prepri{KEK-CP-169 \\  YITP-05-54 \\ KUNS-1987}

\author{
  Hidenori Fukaya, 
  \speaker{Tetsuya Onogi}
  \\
  Yukawa Institute for Theoretical Physics, Kyoto University,
  Kyoto 606-8502, Japan.
}

\author{
  Shoji Hashimoto\\
  High Energy Accelerator Research Organization (KEK),
  Tsukuba 305-0801, Japan.\\
  School of High Energy Accelerator Science,
  The Graduate University for Advanced Studies (Sokendai),
  Tsukuba 305-0801, Japan.
}

\author{
  Takuya Hirohashi\\
  Department of Physics, Kyoto University,
  Kyoto 606-8502, Japan.
}

\author{
  Kenji Ogawa\\
  School of High Energy Accelerator Science,
  The Graduate University for Advanced Studies (Sokendai),
  Tsukuba 305-0801, Japan.
}

\abstract{
The topology conserving gauge action proposed by L\"uscher is
expected to reduce the number of non-smooth gauge configurations 
as well as the topology change compared to the conventional actions 
for the same lattice spacings. 
We report our quenched QCD study of the topological stability
and the scaling violation of the static quark potential. 
We find that the the topology change is indeed suppressed
when the parameter $\epsilon$ is of order one.
We also find that the scaling violation in the static quark
potential remain reasonably small in the parameter range of
our study. 
Our study is done at the inverse lattice spacing
$a^{-1}$ = 1.4--2.5~GeV with the lattice size $L$ =
1.0--1.6~fm. 
}

\FullConference{XXIIIrd International Symposium on Lattice Field Theory\\
                 25-30 July 2005\\
                 Trinity College, Dublin, Ireland}

\begin{document}

\section{Introduction}

The admissibility condition of the gauge fields on the lattice  
\cite{Luscher:1998du,Hernandez:1998et}
\begin{equation}
  \label{eq:admi}
  ||1-P_{\mu\nu}(x)|| < \epsilon \;\;\;\;
  \mbox{for all $x$, $\mu$, $\nu$},
\end{equation}
guarantees the locality and the uniqueness of the index
of the overlap-Dirac operator.
\cite{Neuberger:1997fp,Neuberger:1998wv}.
As an example, the gauge action which restricts the gauge field to 
satisfy the bound (\ref{eq:admi}) 
\begin{equation}\label{eq:admiaction}
  S_G = \beta\sum_{P}\frac{1-\mbox{ReTr}P_{\mu\nu}(x)/3}
  {1-(1-\mbox{ReTr}P_{\mu\nu}(x)/3)/\epsilon},
\end{equation} 
was proposed by L\"uescher \cite{Luscher:1998du}. 
This action may be useful for QCD simulations in two
reasons: 
(1) it may serve to efficiently collect gauge configurations  
with fixed topology $Q\neq 0$ in the $\epsilon$-regime, 
(2) it may give a possibility to reduce the numerical cost 
of dynamical overlap fermions with fewer low-lying modes of
$H_W$ and less frequent topology change.
In fact, this action has been proven to be useful for the 
massive Schwinger model for stabilizing the topological charge
and for improving the chiral behavior of the domain wall
fermions \cite{Fukaya:2003ph,Fukaya:2004kp}.
Also in the four-dimensional quenched QCD, the good stability of the
topological charge has been observed with some reasonable choices of
parameters 
\cite{Shcheredin:2004xa,Bietenholz:2004mq,Shcheredin:2005ma}.  
See also \cite{Nagai}.

In this report we present our quenched study of the topology conserving 
gauge action (\ref{eq:admiaction}) on the stability of the topology
and the scaling. 
Studies on the low-mode distribution in quenched and
dynamical QCD are reported by Matsufuru~\cite{Matsufuru}.

\section{Lattice setup}\label{sec:simulation}

In our study, we take three values of $1/\epsilon$ (= 0, 2/3,
and 1) with three lattice sizes $12^4$, $16^4$, and $20^4$
and four lattice spacings in the range $a^{-1}$ = 1.4--2.5~GeV.
Here,  $1/\epsilon=0$ corresponds to the conventional
plaquette action and $1/\epsilon$ = 2/3 is the boundary,
below which, all gauge configurations are allowed when the
gauge group is $SU(3)$. 
Link variables are generated by the standard hybrid Monte Carlo
algorithm with $\Delta \tau$ = 0.01--0.02 and
$N_{mds}$ = 20--40, 
where $N_{mds}$ is the number of molecular dynamics steps and 
$\Delta \tau$ denotes its step-size. 
We accumulated at least 2,000 trajectories for
thermalization before measuring observables. 
We monitored the plaquette values 
of the gauge fields, but we did observe any case where the
admissibility condition is violated through the hybrid Monte
Carlo updates.

\section{Stability of the topological charge}\label{sec:Qstab}

In order to measure the topological charge, we developed a new 
cooling method, in which we carry out the hybrid Monte Carlo 
simulation using the topology conserving action 
with an exponentially increasing coupling 
$\beta=\beta_{\mathrm{cool}}$ and an exponentially 
decreasing step size $\Delta \tau$ as functions of 
trajectory $n_t$ .
After 50--200 steps, the gauge fields are expected to be 
cooled down close to the classical background in each topological sector.
In fact, the geometrical definition of the topological charge
\cite{Hoek:1986nd}
\begin{equation}\label{eq:topgeo}
Q_{geo}\equiv\frac{1}{32\pi^2}
\sum_{x}\epsilon^{\mu\nu\rho\sigma} \mbox{ReTr}
\left(P_{\mu\nu}(x)P_{\rho\sigma}(x)\right)
\end{equation}
of these ``cooled'' configurations gives numbers close to an
integer.

\begin{figure}[tbp]
  \centering
  \includegraphics[width=8cm]{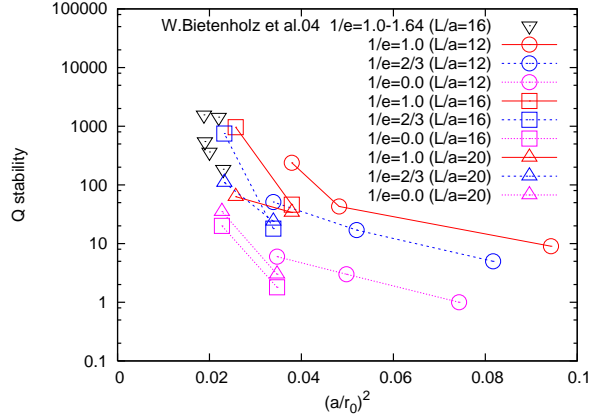}
  \caption{
    Stability of the topological charge
    (\protect\ref{eq:topgeo}) for each parameter set.
    Results are plotted as a function of lattice spacing
    squared.
    Three values of $1/\epsilon$ are distinguished by lines:
    0 (dotted), 2/3 (dashed), 1 (solid).
    Circles, squares, and up-triangles correspond to lattice
    sizes 12, 16 and 20, respectively.
    The down-triangles are from \cite{Bietenholz:2004mq}. 
}
\label{fig:Qstab}
\end{figure}

Figure~\ref{fig:Qstab} shows the stability of the
topological charge defined as
\begin{eqnarray}
\mbox{Stab}_Q \equiv \frac{N_{\mathrm{trj}}}
{\tau_{\mathrm{plaq}}\times \#Q},
\end{eqnarray}
where $\tau_{\mathrm{plaq}}$ is the autocorrelation time
of the plaquette, measured by the method in Appendix E of 
\cite{Luscher:2004rx},
$N_{\mathrm{trj}}$ denotes the number of trajectories and
$\#Q$ is the number of topology changes.
Since the topological charge is measured every 10--20
trajectories, 
$\mbox{Stab}_Q$ gives only an upper limit.
Our results show that the topological charge becomes more
stable for higher $1/\epsilon$, smaller $L$ and smaller $a$.

\section{Static quark potential}\label{sec:scaling}

We calculate the static quark potential $V(\vec{r})$ from 
the Wilson loops $W(\vec{r},t)$ for every 20 trajectories.
The spatial side $\vec{r}$ is taken to be in 6 different
directions of the 3-dimensional unit vectors: 
$\vec{u}$ = (1,0,0), (1,1,0), (2,1,0), (1,1,1), (2,1,1),
(2,2,1). 
We calculate the Sommer scales $r_0$ and $r_c$ 
defined as 
$r_0^2F(r_0)=1.65$  and $r_c^2F(r_c)=0.65$, respectively
\cite{Guagnelli:1998ud,Necco:2001xg}.
$F$ is the force obtained from the static quark
potential. 
We choose a $Q=0$ configuration for the initial condition of
HMC steps and do not take care of the topological charge
hysteresis assuming that the topology of the gauge fields
would not affect the Wilson loops if they are small enough.

\begin{figure}[tbp]
  \centering
  \includegraphics[width=8cm]{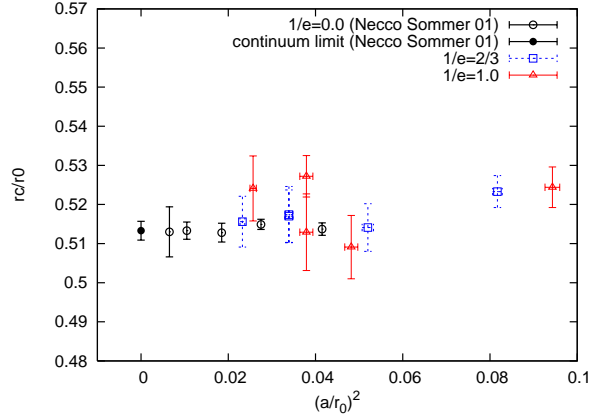}
  \caption{
    Scaling of $r_c/r_0$.
    Squares and triangles are data for $1/\epsilon=2/3$ and
    for 1, respectively.
    Open circles are the result with the plaquette action in
    \cite{Necco:2001xg}, while the filled circle denotes
    their continuum limit.
}
  \label{fig:rcr0}
\end{figure}

\begin{figure}[tbp]
  \centering
  \includegraphics[width=7.5cm]{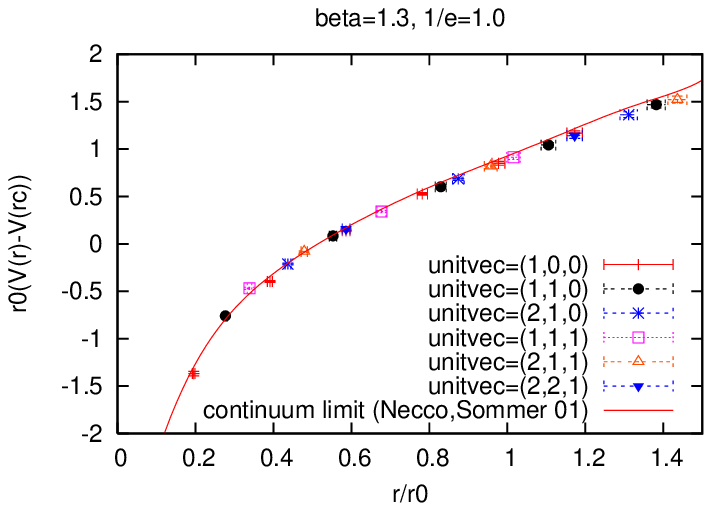}
  \includegraphics[width=7.5cm]{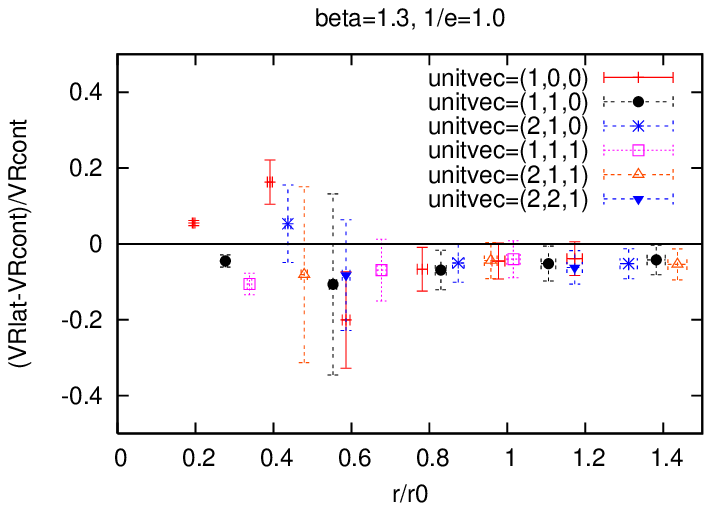}
  \caption{
    Left: heavy quark potential at $\beta=1.3$,
    $1/\epsilon=1.0$ on a $12^4$ lattice. 
    Dashed line shows the continuum limit obtained by an
    interpolation from the result of \cite{Necco:2001xg}.
    Different symbols show the $V(\vec{r})$'s
    with different orientations parallel to $\vec{u}$'s.    
    Right: 
    $(\hat{V}(\vec{r})-\hat{V}_{\mathrm{cont}}(\vec{r}))/
                       \hat{V}_{\mathrm{cont}}(\vec{r})$,
    where $\hat{V}_{\mathrm{cont}}(\vec{r})$ represents the
    continuum limit.
    The error of $\hat{V}_{\mbox{\tiny cont}}(\vec{r})$
    is ignored ($\lesssim 1\%$).
  }
  \label{fig:V(r)}
\end{figure}

Figure~\ref{fig:rcr0} shows the scaling of $r_c/r_0$.
One can see that they agree well with the results with 
the plaquette action and its continuum limit 
\cite{Necco:2001xg} in the region 
$(a/r_0)^2 \lesssim 0.08$.
Also, as seen in Figure~\ref{fig:V(r)} for long distances,
the quark potential itself
$\hat{V}(\vec{r})=r_0(V(\vec{r})-V(r_c))$
does show a good agreement with that in 
the continuum limit obtained with the plaquette action
\cite{Necco:2001xg}. 
Here, $V(r_c)$ is measured using an interpolation polynomial
of order 5. 
For short distances, they show 10--20\% deviations 
due to the violation of the rotational symmetry but the discrepancies
are comparable to those with the plaquette action. 

We conclude that the heavy quark potential for the
admissible gauge fields is quite similar with that of the 
plaquette action and no serious inconsistency can be found.

\section{Perturbative renormalization of the coupling}

\begin{figure}[tbp]
  \centering
  \includegraphics[width=8cm]{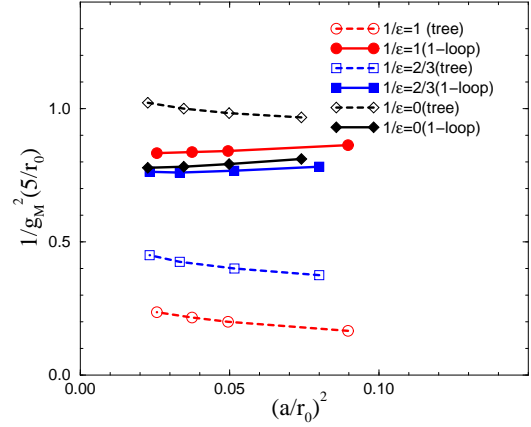}
  \caption{
    $1/g_M^2(5/r_0)$ in the Manton scheme as a function of
    lattice spacing squared.
    Results for various $1/\epsilon$ and $r_0$ are plotted. 
    The bare coupling $1/g_0^2$ is also shown by open
    symbols for comparison.
  }
\label{fig:manton}
\end{figure}

Ellis and Martinelli computed the two-loop corrections to
the gauge coupling for general gauge actions which can be
written by the plaquette \cite{Ellis:1983af}.
Using their formula, the renormalized gauge 
couping $g_{M}$ in the so-called Manton scheme is expressed 
by the bare coupling $g_0$ as
\begin{eqnarray}
\label{eq:Ellis}
\frac{1}{g_M^2(1/a)}
= \frac{1}{g_0^2} + A_1 + A_2 g_0^2,
\end{eqnarray}
where $(A_1,A_2)$ = 
($-$0.2083,$-$0.03056),
(0.34722,$-$0.04783), and
(0.625,$-$0.10276) for $1/\epsilon$ = 0, 2/3, and 1, 
respectively.
In Figure~\ref{fig:manton} we plot the inverse squared coupling
in the Manton scheme at a reference scale $\mu \equiv 5/r_0$ 
as a function of the lattice spacing. 
Here, we use the one-loop formula in (\ref{eq:Ellis}) for
the change of scheme and two-loop renormalization group
equation for the evolution to the reference scale. 
We find that the renormalized coupling for various values of
$1/\epsilon$ shows a good agreement,
which means that the change of $1/\epsilon$ is well
described by the perturbation theory.
One can also see that the scaling violation is small for the
renormalized coupling.
However, we note that the two-loop result gives larger
corrections for larger $1/\epsilon$ so that the convergence
of the perturbative series are poor for $1/\epsilon=1$ and
marginal for $1/\epsilon=2/3$.

\section{Summary}\label{sec:conclusion}

We studied the topology conserving gauge action in the
quenched approximation for various values of $\beta$,
$1/\epsilon$ and the lattice size. 
By measuring the topological charge with a new cooling
method, we find that the stability of the topological charge is
improved for larger $1/\epsilon$ when compared at a same
lattice spacing and lattice size.  
Measuring also the static quark potential and the 
Sommer scales $r_0$, $r_c$, the scaling violation 
are found to be small. 
Our study shows that the topology conserving gauge action 
is feasible for QCD simulations. Practical applications 
of this action such as QCD in $\epsilon$-regime and 
simulations with dynamical overlap fermions are underway.

\section*{ACKNOWLEDGMENTS}\label{sec:acknowlegments}

We would like to thank W.~Bietenholz, 
L.~Del~Debbio, L.~Giusti, M.~Hamanaka, K.~Jansen,
H.~Kajiura, Y.~Kikukawa, M.~L\"uscher, H.~Matsufuru, 
S.~Shcheredin, H.~Suzuki and T.~Umeda for discussions.
The numerical simulation was mainly carried out on the
NEC SX-5 at Research Center for Nuclear Physics.
We also acknowledge the Yukawa Institute for Theoretical
Physics at Kyoto University, where this work was benefited
from the discussions during the YITP-W-04-02 workshop on
"LFT mini workshop''. 
T.~O. is supported by the Grant-in-Aid for Scientific
research from the Ministry of Education, Culture, Sports,
Science and Technology of Japan 
(Nos. 13135213, 16028210, 16540243). 


\end{document}